\documentclass[11pt,prd,preprintnumbers,nofootinbib,superscriptaddress,notitlepage]{revtex4-1}

\usepackage{graphics}
\usepackage{bm}
\usepackage{cancel}
\usepackage{color}
\usepackage{xcolor}
\usepackage{multirow}
\usepackage{slashed}
\usepackage{array}
\usepackage{amssymb}
\usepackage{graphicx}
\usepackage{mathrsfs}
\usepackage{diagbox}
\usepackage{amsmath}
\usepackage{float}
\usepackage{extarrows}
\usepackage[colorlinks,citecolor=blue,anchorcolor=red,menucolor=red,linkcolor=red,filecolor=red,runcolor=red,urlcolor=red,frenchlinks=red]{hyperref}
\usepackage{subfigure}

\makeatletter
\newcommand{\thickhline}{\noalign {\ifnum 0=`}\fi \hrule height 1pt\futurelet \reserved@a \@xhline}
\newcolumntype{"}{@{\hskip\tabcolsep\vrule width 1pt\hskip\tabcolsep}}                             
\makeatother

\def\nn{\nonumber}

\def\gev{\rm GeV}

\def\ev{\rm eV}

\def\mn{m_N^{}}

\def\pslash{\not{\hbox{\kern-4pt $p$}}}
\def\qslash{\not{\hbox{\kern-4pt $q$}}}
\def\lv{\not{\hbox{\kern-4pt $L$}}}
\def\lsim{\mathrel{\raise.3ex\hbox{$<$\kern-.75em\lower1ex\hbox{$\sim$}}}}
\def\gsim{\mathrel{\raise.3ex\hbox{$>$\kern-.75em\lower1ex\hbox{$\sim$}}}}
\def\ifmath#1{\relax\ifmmode #1\else $#1$\fi}

\def\slash{\!\!\!/}

\begin{document}

\title{CP violation in top quark decay via heavy Majorana neutrinos at the LHC}

\author{Peng-Cheng Lu}
\email{pclu@sdu.edu.cn}
\affiliation{School of Physics, Shandong University, Jinan, Shandong 250100, China}

\author{Zong-Guo Si}
\email{zgsi@sdu.edu.cn}
\affiliation{School of Physics, Shandong University, Jinan, Shandong 250100, China}

\author{Zhe Wang}
\email{wzhe@mail.sdu.edu.cn}
\affiliation{School of Physics, Shandong University, Jinan, Shandong 250100, China}

\author{Xing-Hua Yang}
\email{yangxinghua@sdut.edu.cn}
\affiliation{School of Physics and Optoelectronic Engineering, Shandong University of Technology, Zibo, Shandong 255000, China}


\begin{abstract}
The tiny neutrino masses can be explained naturally by extending the standard model with right-handed Majorana neutrinos in low-scale seesaw mechanism,
and the CP violation effect induced by Majorana phase is interesting and worth studying at colliders.
In this paper, we investigate the prospects for measuring CP violation in top quark pair production and rare decay via Majorana neutrinos at the LHC.
The CP asymmetry stems from the significant interference of contributions from two different Majorana neutrinos and can be appreciable when the two neutrino masses are nearly-degenerate.
It is found that in the Majorana neutrino mass range of $10~\gev < \mn < 80~\gev$, the CP asymmetry is independent of the Majorana neutrino mass at the LHC. Any possible new observation of CP violation will be the clear evidence of new physics beyond the standard model.
\end{abstract}



\maketitle


\section{INTRODUCTION}\label{sec1}

The present evidence of neutrino masses and mixings implies the existence of new physics beyond the standard model~\cite{ParticleDataGroup:2020ssz}.
To generate the tiny neutrino masses, the simplest way is to extend the standard model by introducing $n$ right-handed Majorana neutrinos $N_{aR}$ ($a = 1, 2, \cdots , n$).
Since only two active neutrino mass-squared differences have been measured experimentally, the number of right-handed neutrinos is limited to $n\geq2$.
With right-handed Majorana neutrinos, the Dirac neutrino mass terms can be generated through Yukawa interaction after spontaneous gauge symmetry breaking, just like other standard model fermions.
In addition, the introduced Majorana neutrinos and their charge-conjugate counterparts can also form Majorana mass terms as they are $SU(2)_{L}$ gauge singlets.
This is known as the famous type-I seesaw mechanism~\cite{Minkowski:1977sc,Yanagida:1979ss,Gell-Mann:1979ss,Glashow:1979ss,Mohapatra:1979ia}, where the light neutrino masses are inversely proportional to the large mass scale of Majorana neutrinos.
Unfortunately, in the canonical, high-scale type-I seesaw mechanism, the heavy Majorana neutrinos are too heavy and their mixing with active neutrinos is too weak for them to be observed at current and future experiments.
However, there also exists some low-scale type I seesaw mechanisms with the light neutrino masses being directly proportional to a small lepton-number breaking scale.
Two of the most popular scenarios are inverse seesaw~\cite{Mohapatra:1986aw,Mohapatra:1986bd,Bernabeu:1987gr,Gavela:2009cd} and linear seesaw~\cite{Akhmedov:1995ip,Akhmedov:1995vm}.
Moreover, the right-handed neutrinos with masses below the electro-weak scale could also explain the baryon asymmetry of the universe via leptogenesis~\cite{Fukugita:1986hr} and be a natural dark matter candidate~\cite{Dodelson:1993je,Shi:1998km,Abazajian:2001nj,Asaka:2005an}.
From the experimental point of view, we will focus on the low-scale seesaw scenarios.

The existence of Majorana neutrinos will at the same time lead to the violation of lepton number ($\Delta L =2$), which provides an unique opportunity to search for the Majorana neutrinos via lepton-number-violating processes.
For example, the widely studied processes in low energy regions include the well-known neutrinoless double beta decays ($0\nu\beta\beta$)~\cite{Furry:1939qr,Doi:1985dx,Elliott:2004hr}, the rare meson decays~\cite{Abad:1984gh,Littenberg:1991ek,Dib:2000wm,Ali:2001gsa} and the tau lepton decays~\cite{Ilakovac:1995km,Ilakovac:1995wc,Gribanov:2001vv}.
In addition, the same signals can also be searched for at various collider experiments~\cite{Keung:1983uu,Datta:1993nm,Han:2006ip,Si:2008jd,Atre:2009rg,FileviezPerez:2009hdc,Dev:2013wba,Alva:2014gxa,Degrande:2016aje,Antusch:2016ejd,Cai:2017mow,Liu:2019qfa}.
The key point of the lepton-number-violating processes mentioned above is a $W$ decay via Majorana neutrino exchange, which can be specifically expressed as $W^- \rightarrow \ell_{1}^{-}\ell_{2}^{-}(q^\prime\bar{q})^{+}$.
Generally, both the light Majorana neutrinos and the heavy Majorana neutrinos can contribute to the $W$ decay amplitude. However, the light Majorana neutrino contributions are strongly suppressed and can be safely neglected due to their small neutrino masses which are at most ${\cal O}(\ev)$~\cite{Atre:2009rg}.
The CP violation effect will be the smoking gun for new physics beyond the standard model.
If more than one heavy Majorana neutrinos participate in the lepton-number-violating $W$ decay, then it may be possible to produce new sources of CP violation, which can play an important role in explaining the baryon asymmetry of the universe.
To be specific, the CP  asymmetry can be generated from the significant interference of contributions from different Majorana neutrinos.
There have been a great number of theoretical works on examining the prospects for observing CP violation via lepton-number-violating processes, where the CP violation can be measured in the decays of mesons~\cite{Cvetic:2013eza,Cvetic:2014nla,Dib:2014pga,Cvetic:2015naa,Cvetic:2015ura,Cvetic:2020lyh,Godbole:2020doo,Zhang:2020hwj} and tau leptons~\cite{Zamora-Saa:2016ito,Tapia:2019coy}.
Recently, the possibility for measuring CP violation in $W^+$ and $W^-$ decay at the LHC is explored~\cite{Najafi:2020dkp}. However, the CP violation effect is influenced by the initial parton distribution functions (PDF), which leads to the different production cross section of $W^+$ and $W^-$.
To avoid the pollution from PDF, in this paper, we investigate the prospects for measuring CP violation in top quark pair production and rare decay at the LHC, where the number of $W^-$ and $W^+$ bosons coming from the decay of $\bar{t}$ and $t$ is exactly equal.
Since the LHC is a top rich environment, this provides a great opportunity for using the copious $t\bar{t}$ events to investigate the CP violation in top quark decay.
For simplicity, we only consider two heavy Majorana neutrinos, while the general situation with more heavy Majorana neutrinos can be analyzed in a similar way.

This paper is organized as follows.  A phenomenological heavy Majorana model is briefly introduced in Section~\ref{sec2}. The decay of top quark via two different Majorana neutrinos are discussed in Section~\ref{sec3}. Section~\ref{sec4} is devoted to the numerical results and discussions at the LHC. Finally, a short summary is given.

\section{Heavy Majorana neutrino model}\label{sec2}

Following the same notation in Ref.~\cite{Atre:2009rg}, with two right-handed Majorana neutrinos, the mixing relations between the neutrino flavor eigenstates and mass eigenstates can be given by
\begin{equation}
\label{1}
 \nu_{\ell L} = \sum_{m=1}^{3} V_{\ell m} \nu_{m L} + \sum_{m'=1}^{2} R_{\ell m^\prime} N^{c}_{m' L} \; ,
\end{equation}
where $\ell= e, \mu, \tau$ and $ VV^\dag + RR^\dag  = I $.
In terms of the mass eigenstates, the weak charged-current interaction Lagrangian can be written as
\begin{align}
\label{2}
-\mathcal{L}_{\rm cc}
= \frac{g}{\sqrt{2}} W^{+}_{\mu} \sum_{\ell=e}^{\tau} \sum_{m=1}^{3} V_{\ell m}^{\ast} \overline{\nu_{m}} \gamma^{\mu} P_{L} \ell
+ \frac{g}{\sqrt{2}} W^{+}_{\mu} \sum_{\ell=e}^{\tau} \sum_{m'=1}^{2} R_{\ell m^\prime}^{\ast} \overline{N^{c}_{m^\prime}} \gamma^{\mu} P_{L} \ell + \text{h.c.} \; .
\end{align}
Here, $V_{\ell m}$ is the light neutrino mixing matrix that can be measured from the oscillation experiments.
As mentioned above, the contributions of the light Majorana neutrinos to the lepton-number-violating processes can be neglected due to their small masses.
$R_{\ell m^\prime}$ indicates the mixing between charged-leptons and heavy Majorana neutrinos and can be parameterized as~\cite{Xing:2007zj}
\begin{equation}
\label{3}
R_{\ell m^\prime}=\left|R_{\ell m^\prime}\right|e^{i\phi_{\ell m^\prime}} , ~~~ \ell= e, \mu, \tau, ~~~ m^\prime=1, 2 \; .
\end{equation}
The complex phase $\phi_{\ell m^\prime}$ can serve as one of the new sources of CP violation, which can be determined from possible collider experiments.

The masses of the heavy Majorana neutrinos $m_{N}$ and the mixing elements $R_{\ell m^\prime}$ are strongly restricted by experimental observations (for review, we refer to Ref.~\cite{Deppisch:2015qwa}).
So far, the most stringent bound on mixing with electrons can be derived from $0\nu\beta\beta$-decay experiments~\cite{GERDA:2018pmc}:
\begin{eqnarray}
\label{4}
\sum_{i} \frac{\left|R_{ei}\right|^2}{m_{N_{i}}} < 5 \times 10^{-5}~{\rm TeV^{-1}} \; .
\end{eqnarray}
It is worth mentioning that the $0\nu\beta\beta$ constraint is usually model dependent, and may be significantly weakened in certain cases~\cite{Pascoli:2013fiz}.
For $m_N < m_Z$, 95\% C.L. limits on the mixing parameters $R_{\ell m^\prime}$ can be obtained from a reanalysis of the LEP data~\cite{DELPHI:1996qcc}:
\begin{eqnarray}
\label{5}
\left|R_{e1}\right|^2, ~ \left|R_{\mu 1}\right|^2, ~ \left|R_{\tau 1}\right|^2 < {\cal O} (10^{-5}) \; .
\end{eqnarray}
A global fit to lepton flavor and electro-weak precision data has been performed to constrain the size of $R_{\ell m^\prime}$ for heavy neutrino mass above the electro-weak scale~\cite{Fernandez-Martinez:2016lgt}.
At 95\% C.L., the limits are
\begin{eqnarray}
\label{6}
\sum_{i} \left|R_{e i}\right|^2 < 2.5 \times 10^{-3} \; ,~~  \sum_{i} \left|R_{\mu i}\right|^2 < 4.4 \times 10^{-4} \; , ~~ \sum_{i} \left|R_{\tau i}\right|^2 < 5.6 \times 10^{-3} \; .
\end{eqnarray}
At the LHC experiments, the most restrictive direct limits on the mixing parameters $|R_{e1}|^2$ and $|R_{\mu 1}|^2$ for heavy Majorana neutrino mass between 20 GeV and 1600 GeV are varying from $2.3 \times 10^{-5}$ to unity~\cite{CMS:2018jxx}.
In this paper, we will adopt a phenomenological approach where the masses and mixing parameters of the heavy neutrinos are simply parameterized as free parameters.
To be conservative, the values of the mixing parameters are set as
\begin{eqnarray}
\label{7}
\left|R_{e i}\right|^2 = 1.0 \times 10^{-7}  \; ,~~ \left|R_{\mu i}\right|^2 = \left|R_{\tau i}\right|^2 = 1.0 \times 10^{-5} \; , \; {\rm for} ~~ i = 1, 2 \; .
\end{eqnarray}

\section{CP violation in top (anti-top) quark decay}\label{sec3}

Given the charged-current interaction Lagrangian in Eq.~(\ref{2}), the decay of top and anti-top quark with $\Delta L=2$ via heavy Majorana neutrinos can be respectively expressed by
\begin{align}
\label{8}
t(p_1) &\to b(p_2) + l_{\alpha}^{+}(p_3) + N_{i}(p_N) \to b(p_2) + l_{\alpha}^{+}(p_3) + l_{\beta}^{+}(p_4) + q(p_5) + \bar{q}^\prime(p_6) \; , \nn \\
\bar{t}(p_1) &\to \bar{b}(p_2) + l_{\alpha}^{-}(p_3) + N_{i}(p_N) \to \bar{b}(p_2) + l_{\alpha}^{-}(p_3) + l_{\beta}^{-}(p_4) + \bar{q}(p_5) + q^\prime(p_6) \; ,
\end{align}
where $\alpha, \beta = e, \mu, \tau$ and $ i = 1, 2 $. $p_1$, $p_2$ etc. represent the 4-momentum of the corresponding particles.
The differential decay width for the process in Eq.~(\ref{8}) can be given by
\begin{eqnarray}
\label{8.1}
d\Gamma \left(t \to b l_{\alpha}l_{\beta} j_1 j_2\right)=\frac{1}{2m_t}\overline{\left|{\cal M}\right|^2} d{\cal L}ips_5 \; .
\end{eqnarray}
Here, $d{\cal L}ips_5$ represents the five-body Lorentz invariant phase space of the final particles.
$\overline{|{\cal M}|^2}$ is the squared scattering amplitude averaged (summed) over the initial (final) particles, and can be expressed as
\begin{align}
\label{8.2}
\overline{\left|{\cal M}_{\ell_\alpha^\pm\ell_\beta^\pm}\right|^2} =& \frac{g^8}{m_W^4}\left|V_{tb}\right|^2\left|V_{qq^\prime}\right|^2\left(1-\frac{1}{2}\delta_{\alpha\beta}\right)
\left|D_{W}\left(p_w^2\right)\right|^2\left|D_{W}\left({p_w^\prime}^2\right)\right|^2  \nn \\
& \times \left\{m_{N_1}^{2}\left|R_{\alpha 1}R_{\beta 1}\right|^2{\cal T}_1 + m_{N_2}^{2}\left|R_{\alpha 2}R_{\beta 2}\right|^2{\cal T}_2 \right. \nn \\
& \left. + m_{N_1} m_{N_2} \left|R_{\alpha 1}R_{\alpha 2}R_{\beta 1}R_{\beta 2}\right|{\rm Re}\left[e^{\pm i\Delta\phi}{\cal T}_{12}\right] \right\} \; ,
\end{align}
where $p_w=p_1-p_2$, $p_w^\prime=p_5+p_6$. $V_{tb}$ $\left(V_{qq^\prime}\right)$ is the CKM matrix element and $R_{\alpha i }$ ($\alpha = e, \mu, \tau$ and $i = 1, 2$) is the mixing matrix element defined in  Eq.~(\ref{3}). $\Delta\phi=\phi_{\alpha 2} - \phi_{\alpha 1} +  \phi_{\beta 2} -  \phi_{\beta 1}$ is the CP phase difference induced by the
significant interference between $N_1$ and $N_2$. The explicit expressions of $D_W$, ${\cal T}_i$ ($i = 1, 2$) and ${\cal T}_{12}$ are shown in Appendix~\ref{appA}.

With only one heavy Majorana neutrino, the process in Eq.~(\ref{8}) has been well studied in the literature~\cite{Si:2008jd,Liu:2019qfa}.
However, as can be seen in Eq.~(\ref{8.1}) and Eq.~(\ref{8.2}), the squared scattering amplitude $\overline{|{\cal M}|^2}$, and thus the differential decay width $d\Gamma$, of top (anti-top) quark decay is irrelevant to the complex phase $\Delta\phi$, which means that there is no CP violation in this case.
Therefore, in order to generate the CP-violating asymmetry, there must exist at least two different heavy Majorana neutrinos and the CP phase difference $\Delta\phi$ indicated in Eq.~(\ref{8.2}) cannot be zero.
In the subsequent discussions, we generalize the previous works and consider the case of two intermediate on-shell Majorana neutrinos $N_1$ and $N_2$.
The significant interference between the $N_1$ and $N_2$ contributions can lead to a difference in the rates of top quark decay and its CP-conjugate process, which is a signal of CP violation. Furthermore, the CP violation effect can be appreciable when the two neutrino masses are nearly-degenerate.

As demonstrated in Ref.~\cite{Si:2008jd}, in the narrow-width approximation, the total decay width of top quark can be factorized as
\begin{eqnarray}
\label{9}
\Gamma \left(t \to b l_{\alpha}l_{\beta} j_1 j_2\right)
\approx \Gamma \left(t \to b l_{\alpha} N\right)~{\rm Br}\left(N \to l_{\beta} j_1 j_2\right)
= \left(2-\delta_{\alpha\beta}\right)~S_{\alpha\beta}~\Gamma_{0} \; .
\end{eqnarray}
$\Gamma_{0}$ is the reduced decay width, and basically independent of all the mixing parameters.
$S_{\alpha\beta}$ is the so-called ``effective mixing parameter" and can be defined as
\begin{eqnarray}
\label{10}
S_{\alpha\beta}=\frac{\left|R_{\alpha N}R_{\beta N}\right| ^{2}}{\sum_{\ell=e}^{\tau} \left|R_{\ell N}\right| ^{2}} \; .
\end{eqnarray}
In our calculations, we employ the same definition for $S_{\alpha\beta}$ in the case with $N_1$ and $N_2$.
For simplicity, the CKM matrix is considered as diagonal with unit entries, and only the top quark mass are taken into account. Moreover, the following assumptions are applied
\begin{eqnarray}
\label{11}
m_{N_{2}} = m_{N_{1}} + \Gamma_{N_{1}}/2  \; , ~~ \Gamma_{N_{2}} \approx \Gamma_{N_{1}} \; .
\end{eqnarray}

\begin{figure}[!htbp]
\begin{center}
\subfigure[]{\label{fig1a}
\includegraphics[width=0.4\textwidth]{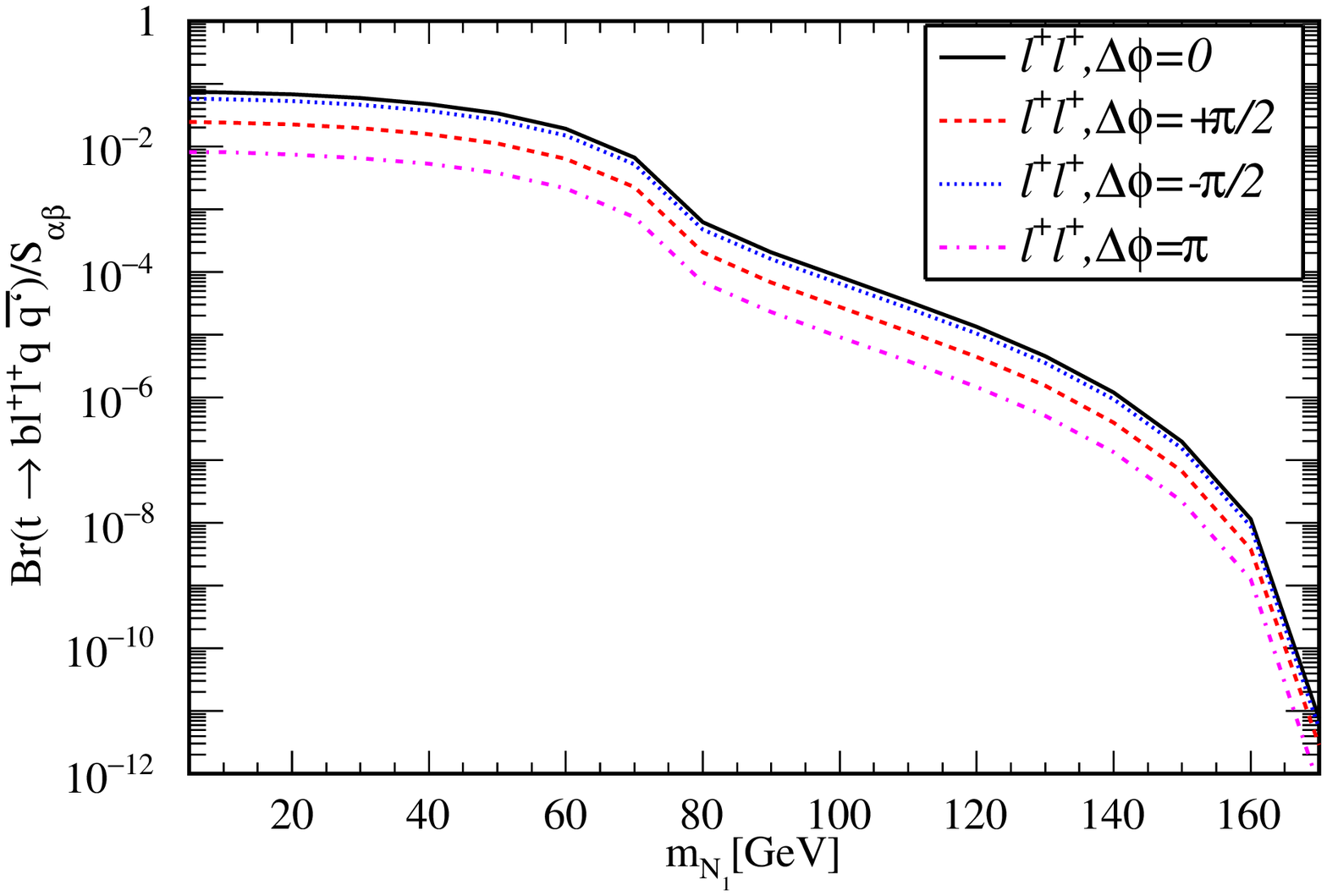} }
\hspace{-0.5cm}~
\subfigure[]{\label{fig1b}
\includegraphics[width=0.4\textwidth]{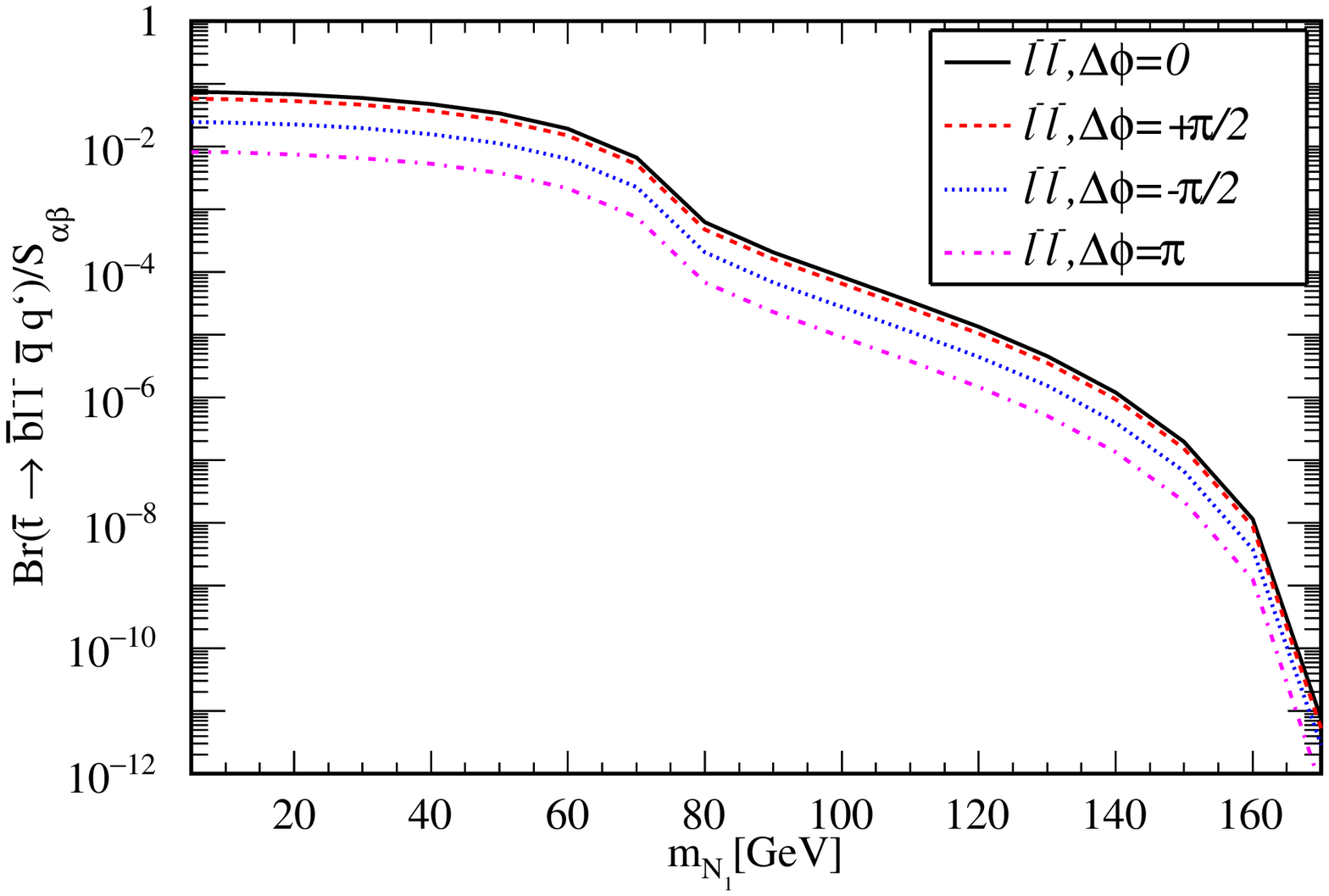} }
\caption{ The normalized branching ratio of (a) top quark decay $t \to b l_{\alpha}^{+}l_{\beta}^{+} q \bar{q}^\prime$ and (b) anti-top quark decay $\bar{t} \to \bar{b} l_{\alpha}^{-}  l_{\beta}^{-}  \bar{q}  q^\prime$ via $N_1$ and $N_2$ for $\Delta\phi=0, +\pi/2, -\pi/2, \pi$.}\label{fig1}
\end{center}
\end{figure}

The normalized branching ratio of top quark decay $t \to b l_{\alpha}^{+}l_{\beta}^{+} q \bar{q}^\prime$ and anti-top quark decay $\bar{t} \to \bar{b} l_{\alpha}^{-}  l_{\beta}^{-}  \bar{q}  q^\prime$ via $N_1$ and $N_2$ are shown in Fig.~\ref{fig1} as a function of $m_{N_{1}}$, where the CP phase difference is set to $\Delta\phi=0, +\pi/2, -\pi/2, \pi$.
From this plot, one can find that, for fixed $\Delta\phi$, as the Majorana neutrino mass increases, the normalized branching ratio decreases.
When $m_{N_{1}}>m_W$, the branching ratio becomes less than $10^{-3}S_{\alpha\beta}$ for $\Delta\phi=0$ and $10^{-4}S_{\alpha\beta}$ for $\Delta\phi=\pi$.
At the low Majorana neutrino mass region, the search for Majorana neutrinos has been studied extensively in heavy meson decays.
In order to effectively measure the CP violation in top quark decay, the Majorana neutrino mass range of our interest is limited to $10~\gev < \mn < 80~\gev$.
Moreover, the branching ratio for $\Delta\phi=+\pi/2~(-\pi/2)~[\pi]$ is just about 60\% (30\%) [10\%] of that for $\Delta\phi=0$ in the case of top quark decay,
but 30\% (60\%) [10\%] in the anti-top quark decay case.

\begin{figure}[!htbp]
\begin{center}
\subfigure[]{\label{fig4.1a}
\includegraphics[width=0.4\textwidth]{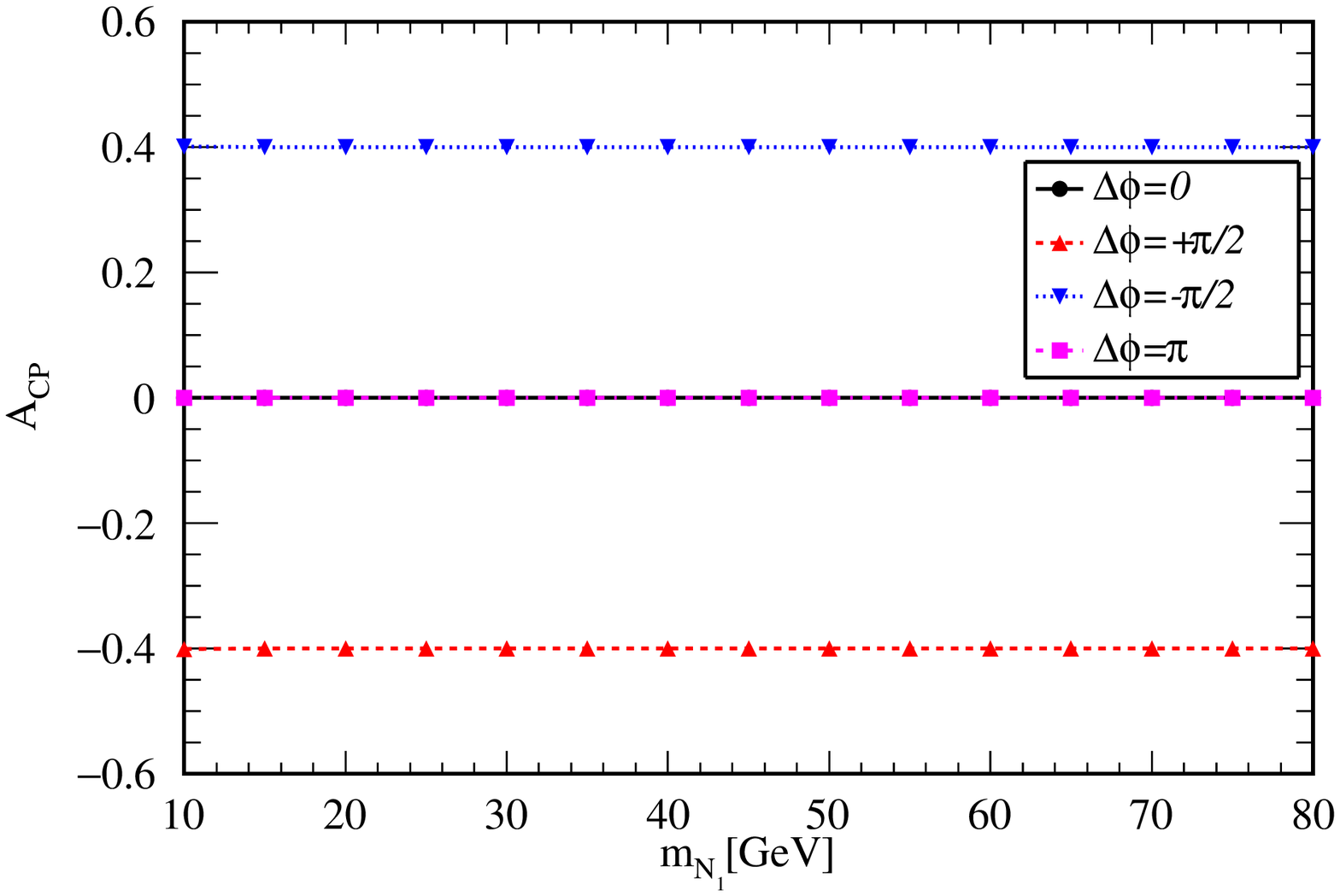} }
\hspace{-0.5cm}~
\subfigure[]{\label{fig4.1b}
\includegraphics[width=0.4\textwidth]{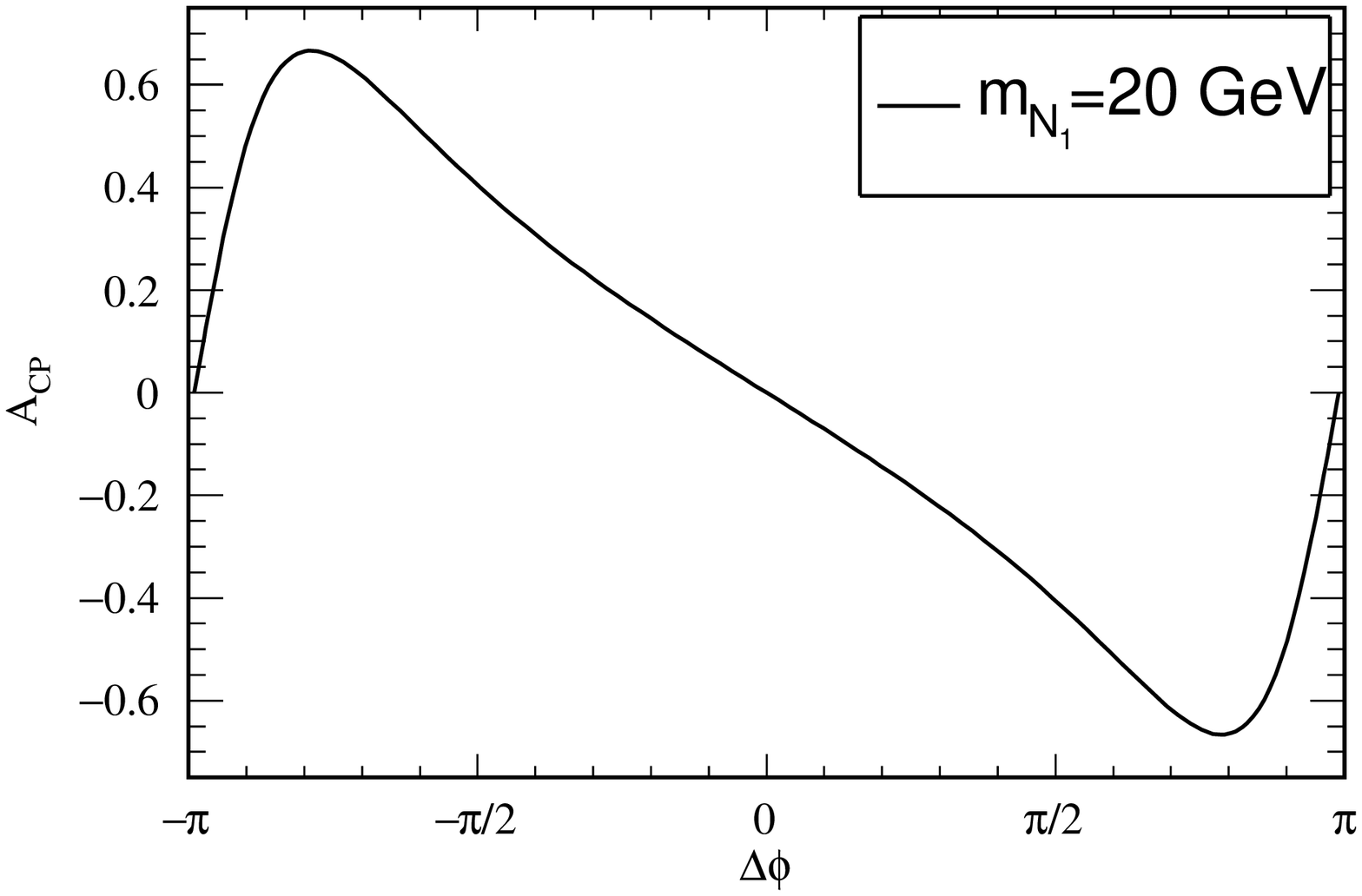} }
\caption{(a) The value of ${\cal A}_{\rm CP}$ as a function of $m_{N_{1}}$ for $\Delta\phi=0, +\pi/2, -\pi/2, \pi$. (b) The value of ${\cal A}_{\rm CP}$ as a function of $\Delta\phi$ for $m_{N_{1}} = 20~\gev$. }\label{fig4.1}
\end{center}
\end{figure}

The difference between the rates of $t \to b l_{\alpha}^{+}l_{\beta}^{+} q \bar{q}^\prime$ and its CP-conjugate process $\bar{t} \to \bar{b} l_{\alpha}^{-}l_{\beta}^{-} \bar{q} q^\prime$ can then induce the CP asymmetry ${\cal A}_{\rm CP}$, which can be defined as
\begin{eqnarray}
\label{15}
{\cal A}_{\rm CP} = \frac{\Gamma\left(t \to b l_{\alpha}^{+}l_{\beta}^{+} q \bar{q}^\prime\right)-\Gamma\left(\bar{t} \to \bar{b} l_{\alpha}^{-}l_{\beta}^{-} \bar{q} q^\prime\right)}
{\Gamma\left(t \to b l_{\alpha}^{+}l_{\beta}^{+} q \bar{q}^\prime\right)+\Gamma\left(\bar{t} \to \bar{b} l_{\alpha}^{-}l_{\beta}^{-} \bar{q} q^\prime\right)} .
\end{eqnarray}
The numerical results of ${\cal A}_{\rm CP}$ as a function of $m_{N_{1}}$ for various values of $\Delta\phi$ are shown in Fig.~\ref{fig4.1a}.
For comparison, we also display the value of ${\cal A}_{\rm CP}$ as a function of $\Delta\phi$ for $m_{N_{1}} = 20~\gev$ in Fig.~\ref{fig4.1b}.
It is found that, for $\Delta\phi=0, \pi$, the CP asymmetry disappears.
Furthermore, we see that the size of ${\cal A}_{\rm CP}$ is basically independent of the Majorana neutrino mass, and ${\cal A}_{\rm CP} \to -{\cal A}_{\rm CP}$ for negative values of $\Delta\phi$.
The maximal values of ${\cal A}_{\rm CP}\approx0.6$ can be reached for $\Delta\phi\approx\pm4\pi/5$.


\section{CP violation in top anti-top pair production and rare decay at the LHC}\label{sec4}

\begin{figure}[!htbp]
\begin{center}
\subfigure[]{\label{fig3a}
\includegraphics[width=0.4\textwidth]{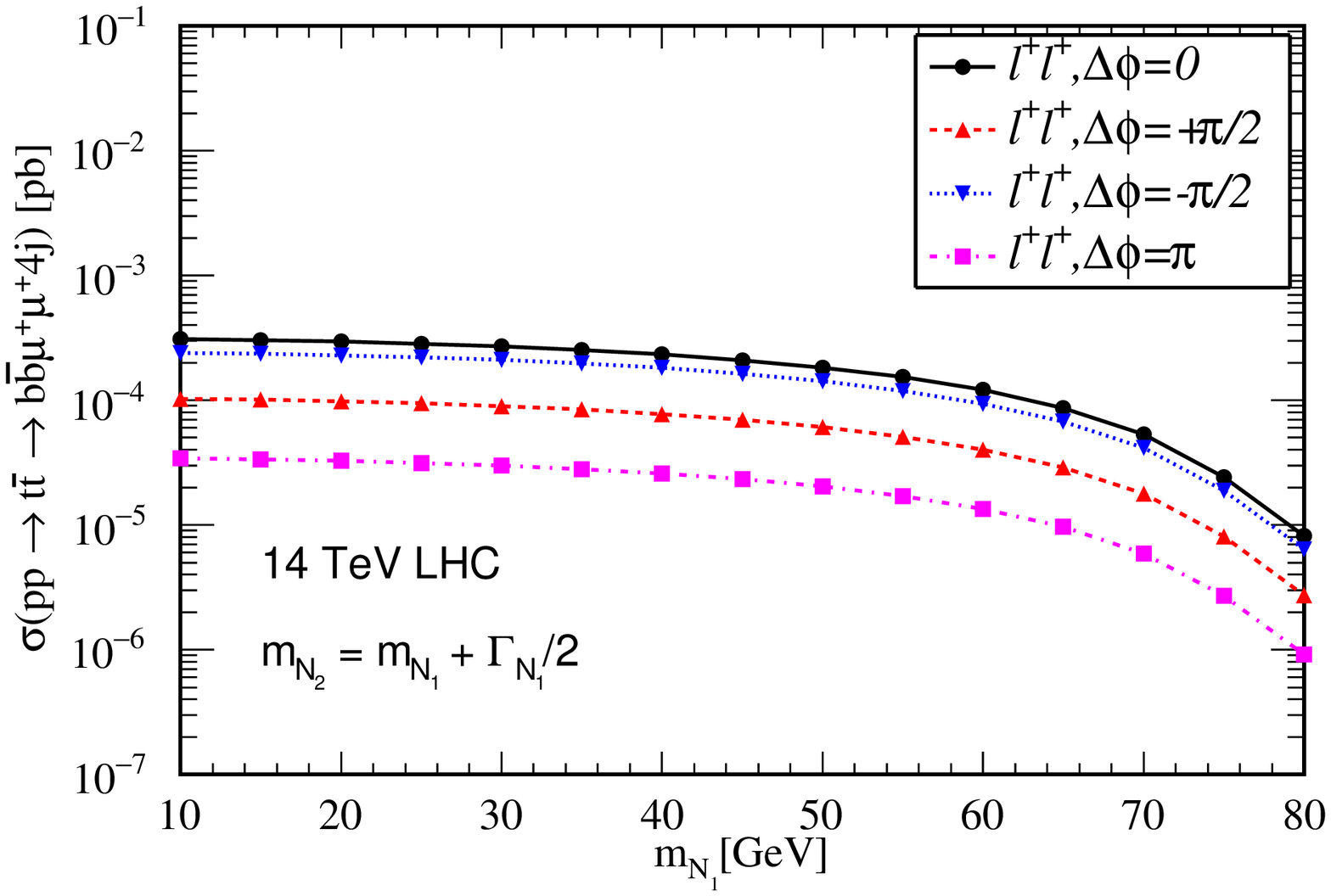} }
\hspace{-0.5cm}~
\subfigure[]{\label{fig3b}
\includegraphics[width=0.4\textwidth]{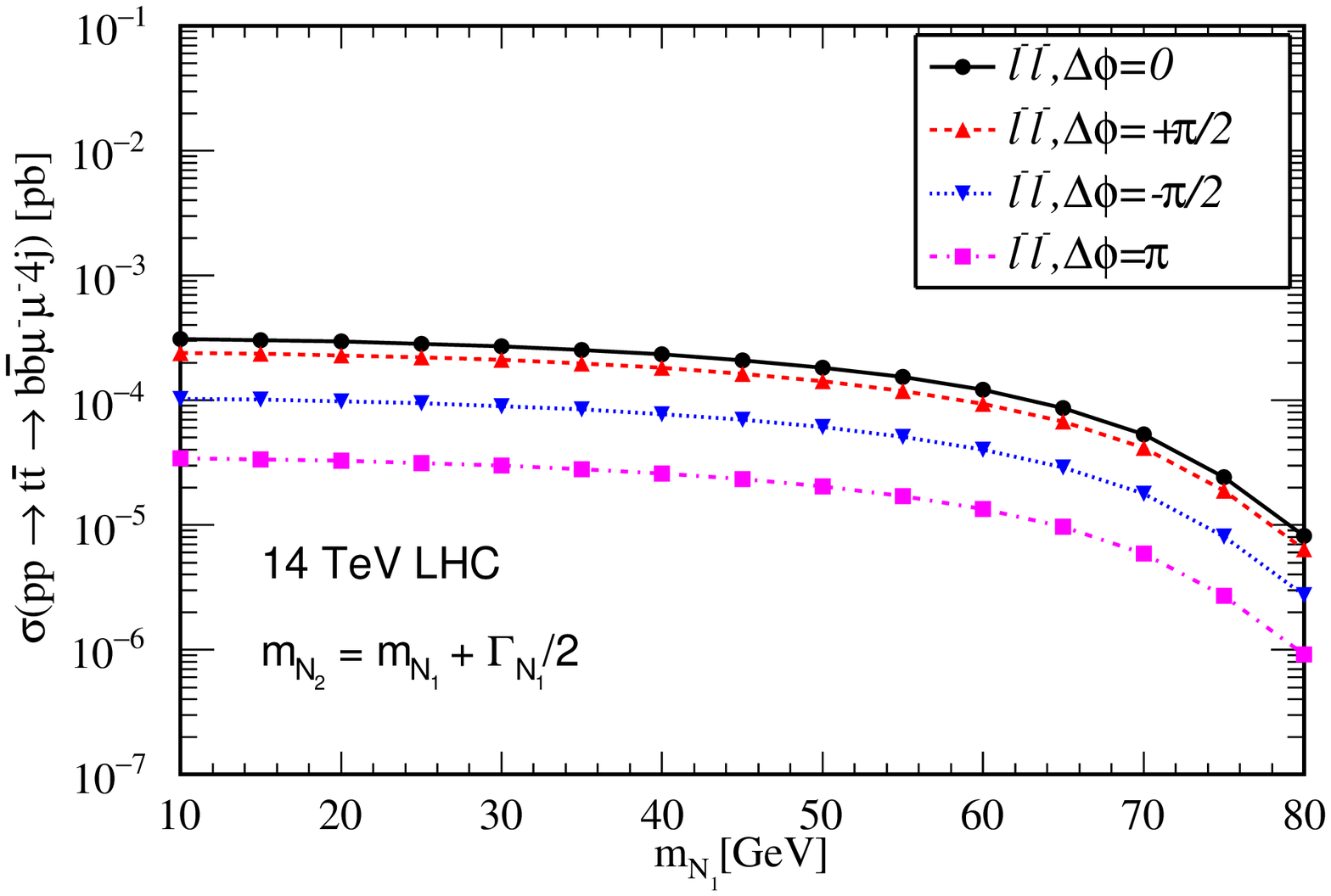} }
\hspace{-0.5cm}~
\subfigure[]{\label{fig3c}
\includegraphics[width=0.4\textwidth]{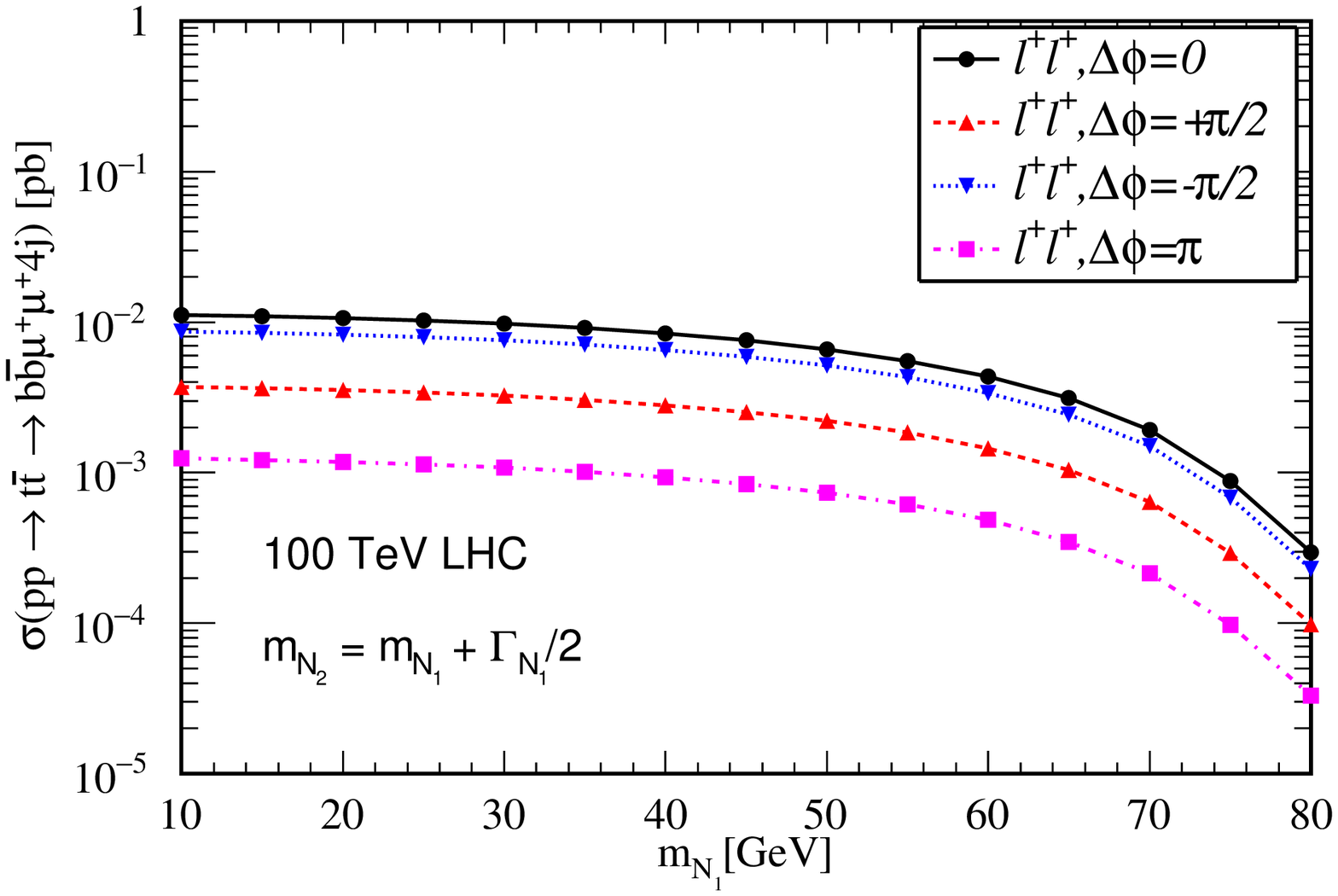} }
\hspace{-0.5cm}~
\subfigure[]{\label{fig3d}
\includegraphics[width=0.4\textwidth]{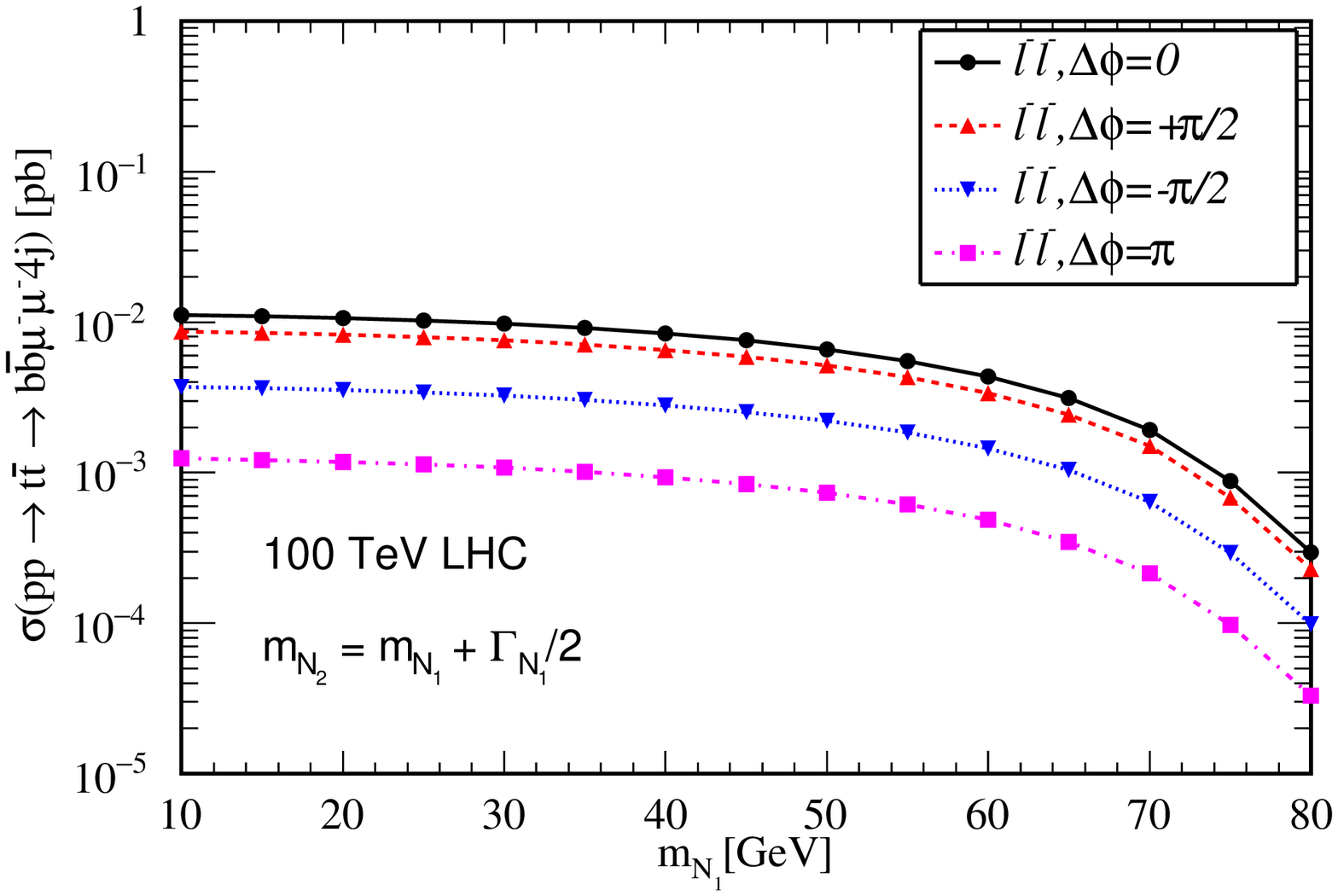} }
\caption{The total cross sections for (a) $pp \to t\bar{t} \to b\bar{b} \mu^+ \mu^+ 4j$ and (b) $pp \to t\bar{t} \to b\bar{b} \mu^- \mu^- 4j$ at 14 TeV LHC, (c) $pp \to t\bar{t} \to b\bar{b} \mu^+ \mu^+ 4j$ and (d) $pp \to t\bar{t} \to b\bar{b} \mu^- \mu^- 4j$ at 100 TeV LHC, as a function of $m_{N_{1}}$. Here the CP phase difference is set to $\Delta\phi=0, +\pi/2, -\pi/2, \pi$. }\label{fig3}
\end{center}
\end{figure}

As a top rich environment, the LHC will offer a great opportunity to precisely explore the prospects for measuring such a CP asymmetry.
At the LHC, the $t\bar{t}$ pairs are dominantly produced by the strong interactions~\cite{Nason:1987xz,Bernreuther:2001rq,Bernreuther:2004jv}.
In this paper, we consider the following process
\begin{eqnarray}
\label{14}
pp \to t\bar{t} \to b\bar{b} + l^\pm l^\pm  + 4j \; .
\end{eqnarray}
where one top quark in leptonic decay and the remain one is required to decay hadronically.
Since the detection of muon leptons is most efficient at the LHC, we will thus concentrate on the clean dimuon production channel.
In Fig.~\ref{fig3}, we display respectively the total cross sections for the same-sign dilepton production process as a function of $m_{N_{1}}$ at 14 TeV and 100 TeV LHC, where the CP phase difference is set to $\Delta\phi=0, +\pi/2, -\pi/2, \pi$.
Here, we employ the CTEQ6L1 for the parton distribution functions in proton~\cite{Pumplin:2002vw}.

\begin{figure}[!htbp]
\begin{center}
\subfigure[]{\label{fig4a}
\includegraphics[width=0.4\textwidth]{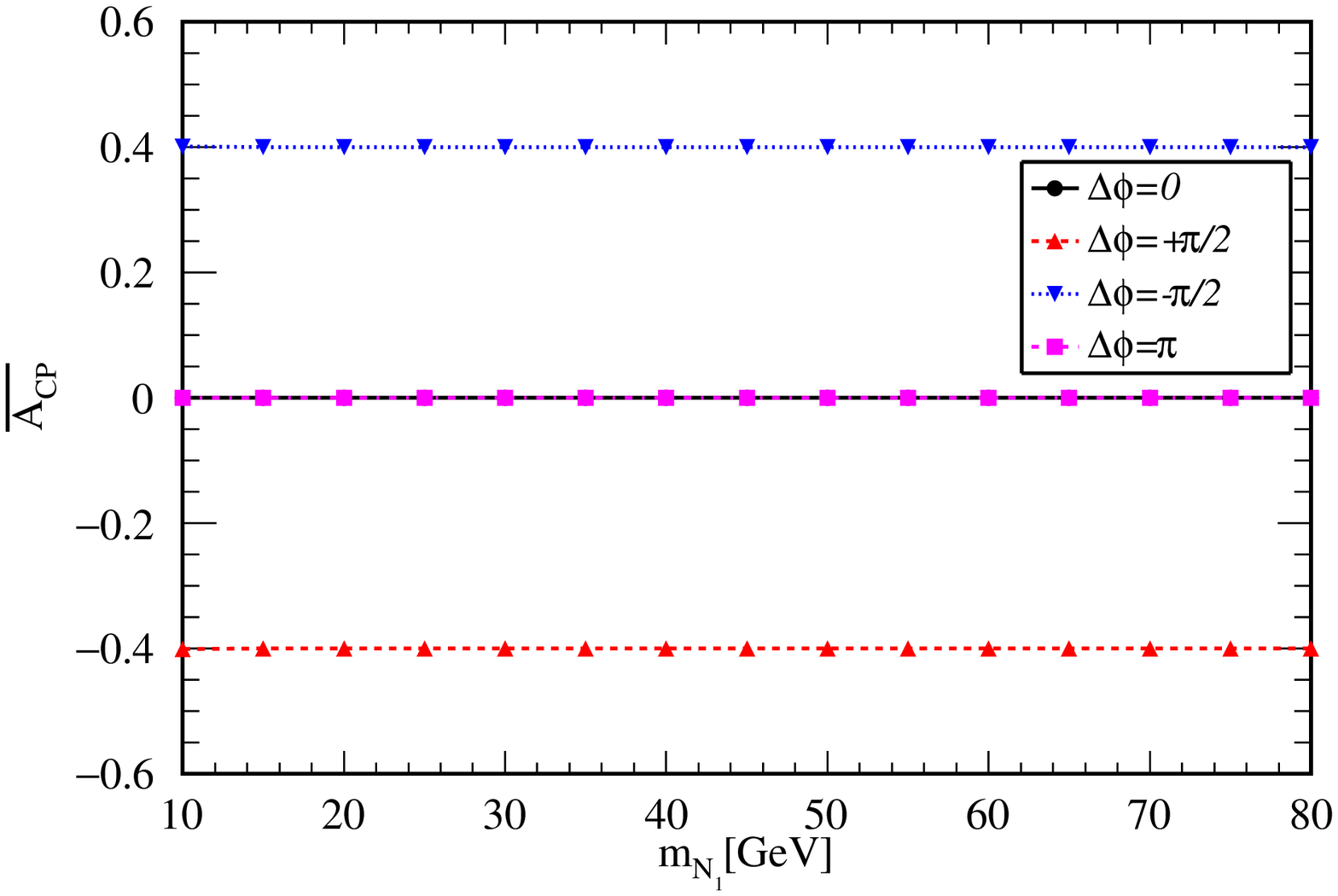} }
\hspace{-0.5cm}~
\subfigure[]{\label{fig4b}
\includegraphics[width=0.4\textwidth]{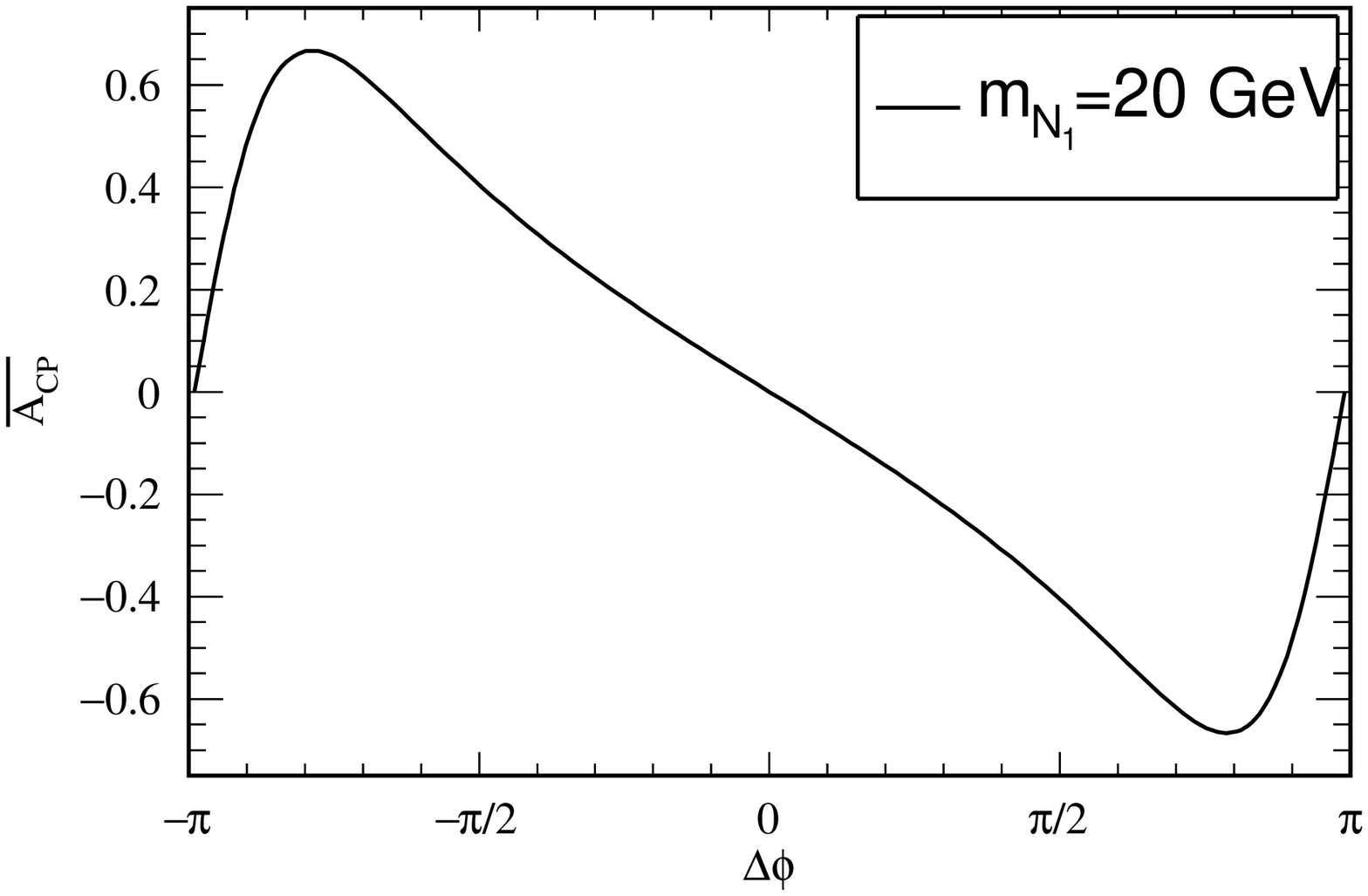} }
\caption{(a) The value of ${\cal A}_{\rm CP}$ as a function of $m_{N_{1}}$ for $\Delta\phi=0, +\pi/2, -\pi/2, \pi$. (b) The value of ${\cal A}_{\rm CP}$ as a function of $\Delta\phi$ for $m_{N_{1}} = 20~\gev$. }\label{fig4}
\end{center}
\end{figure}

To investigate the CP violation at the LHC, the CP asymmetry can be defined as
\begin{eqnarray}
\label{15}
\overline{{\cal A}_{\rm CP}} = \frac{\sigma(pp \to b\bar{b} \mu^+ \mu^+ 4j)-\sigma(pp \to b\bar{b} \mu^- \mu^- 4j)}
{\sigma(pp \to b\bar{b} \mu^+ \mu^+ 4j)+\sigma(pp \to b\bar{b} \mu^- \mu^- 4j)} .
\end{eqnarray}
The numerical results of $\overline{{\cal A}_{\rm CP}}$ are shown in Fig.~\ref{fig4}.
One finds that the tendency of the CP asymmetry $\overline{{\cal A}_{\rm CP}}$ in Fig.~\ref{fig4} is same as that in Fig.~\ref{fig4.1}. This is exactly what we expected.
The reason is that the CP asymmetry only stems from the rate difference between top quark decay and its CP-conjugate process.

Our signal consists of 2 same-sign dilepton plus 6 jets (including 2 b-jets) associated with no significant missing transverse energy.
As the jets originating from the top quark five-body decay $t \to b l_{}^{+}l_{}^{+} j j$ are much softer than those in the decay of $\bar{t} \to \bar{b} j j$, the 6 jets in final state may be merged into 4 jets plus one fat jet.
In our calculations, the number of the final state jets is limited to $n_j=5$.
Though there is no source of missing transverse momentum in our signal process, the missing transverse momentum may appear due to the detector-level mis-measurements.
To simulate the detector effects, we smear the lepton and jet energies according to the assumption of the Gaussian resolution parametrization
\begin{eqnarray}
\label{16}
\frac{\delta(E)}{E} = \frac{a}{\sqrt{E}}\oplus b,
\end{eqnarray}
where $a=5\%$, $b=0.55\%$ for leptons and $a=100\%$, $b=5\%$ for jets, respectively~\cite{CMS:2007sch,ATLAS:2009zsq}.
In order to identify the isolated lepton or jet, the angular separation between particle $i$ and particle $j$ can be defined as
\begin{eqnarray}
\label{17}
\Delta R_{ij} = \sqrt{\Delta \phi^2_{ij}+\Delta\eta^2_{ij}} \; .
\end{eqnarray}
In the following numerical calculations, we apply the basic acceptance cuts (referred as cut-I)
\begin{eqnarray}
\label{18}
p_{T}^{\ell} > 10~{\rm GeV} \; , \; |\eta^{\ell}| < 2.8 \; , \; p_{T}^{j} > 15~{\rm GeV} \; , \; |\eta^{j}| < 3.0 \; , \; 0.4 < \Delta R_{\ell j} < 2.5 \; , \; n_j=5 \; .
\end{eqnarray}
To maximize the contributions to our signal rate, we further demand the missing transverse energy satisfies (referred as cut-II)
\begin{eqnarray}
\label{19}
 {E\slash}_{T} < 20~{\rm GeV} \; .
\end{eqnarray}
The dominant backgrounds in the standard model for our signal process are $pp \to t\bar{t}W^{\pm}j \to b\bar{b}\mu^{\pm}\mu^{\pm}jjj+{E\slash}_{T}$
and $pp \to b\bar{b}W^{\pm}W^{\pm}jjj \to b\bar{b}\mu^{\pm}\mu^{\pm}jjj+{E\slash}_{T}$, which are simulated by \textsf{MadGraph5\_aMC@NLO}~\cite{Alwall:2014hca}.
After analyzing these kinds of backgrounds, we find that the backgrounds $b\bar{b}W^{\pm}W^{\pm}jjj$ are much smaller and can be neglected.
The parton shower and hadronization are performed with \textsf{Pythia-8.2}~\cite{Sjostrand:2006za} and the fast detector simulation are simulated with \textsf{PGS}~\cite{pgs}.
Jets-clustering is done by \textsf{FastJet}~\cite{Cacciari:2011ma} with the anti-$k_t$ algorithm~\cite{Cacciari:2008gp}.

To purify the signal, we employ the procedure described in Ref.~\cite{Si:2008jd} to fully reconstruct the two tops.
First, the 3 jets with invariant mass closest to $m_t$ are selected to reconstruct the hadronic top.
Second, all the remain ingredients are grouped to reconstruct the leptonic top.
The following cut on the top reconstruction is applied (referred as cut-III)
\begin{eqnarray}
\label{20}
|M_{\rm inv}-m_t|< 20~\gev  \; .
\end{eqnarray}

\begin{figure}[!htbp]
\begin{center}
\includegraphics[width=0.4\textwidth]{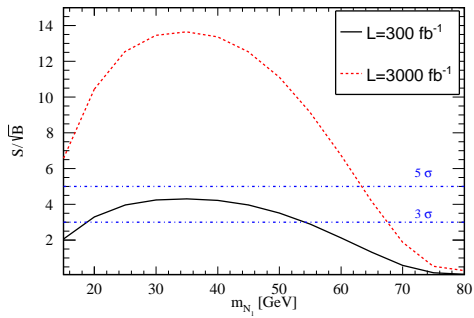}
\caption{The statistical significance $S/\sqrt{B}$ as a function of $m_{N_{1}}$ with the integrated luminosity of $\cal{L}$=300~$\mathrm{fb}^{-1}$ and $\cal{L}$=3000~$\mathrm{fb}^{-1}$ at 14 TeV LHC, where the CP phase difference is set to $\Delta\phi=\pi/2$.}\label{fig5}
\end{center}
\end{figure}

Taking $\Delta\phi=\pi/2$ as an example, with the integrated luminosity of $\cal{L}$=300~$\mathrm{fb}^{-1}$ and $\cal{L}$=3000~$\mathrm{fb}^{-1}$, the statistical significance $S/\sqrt{B}$ as a function of $m_{N_{1}}$ at 14 TeV LHC is displayed in Fig.~\ref{fig5}, where $S$ and $B$ denote the signal and background event numbers after all the cuts in Eqs.~(\ref{18})-(\ref{20}), respectively.
We find that, a 3$\sigma$ discovery can be made for $19~\gev<m_{N_{1}}<54~\gev$ with $\cal{L}$=300~$\mathrm{fb}^{-1}$ at 14 TeV LHC. With $\cal{L}$=3000~$\mathrm{fb}^{-1}$, the heavy Majorana neutrino mass can reach $m_{N_{1}} \simeq 64~\gev$ for 5$\sigma$ discovery. Moreover, at $m_{N_{1}} \simeq 35~\gev$, the statistical significance takes the maximum value $S/\sqrt{B} \simeq 13.5$.
The cross sections for the signal and background processes at 14 TeV and 100 TeV LHC after all cuts are shown in Table~\ref{table1}. Also shown is the statistical significance $S/\sqrt{B}$ with integrated luminosity of $\cal{L}$=300~$\mathrm{fb}^{-1}$ and $\cal{L}$=3000~$\mathrm{fb}^{-1}$.
For illustration, we have used $m_{N_{1}}=30~\gev$ and $\Delta\phi=\pi/2$.
It is shown that after all the cuts, the signal cross section remains only $4.35\times 10^{-3}$~$\mathrm{fb}^{-1}$ at 14 TeV LHC, which is hard to be detected. However, it is possible to do the study at 100 TeV LHC, and the corresponding statistical significance can reach 37.18 (117.60) with $\cal{L}$=300~$\mathrm{fb}^{-1}$ ($\cal{L}$=3000~$\mathrm{fb}^{-1}$).
Finally, we show the value of $\overline{{\cal A}_{\rm CP}}$ at 14 TeV and 100 TeV LHC after all cuts in Table~\ref{table2}, where $m_{N_1}$ =30 GeV and $\Delta\phi=\pm\pi/4, \pm\pi/2, \pm4\pi/5$ are used for illustration.
The CP violation effect can be used as the evidence for new physics beyond the standard model.

\begin{table}[!htbp]
  \caption{The cross sections for the signal and background processes at 14 TeV and 100 TeV LHC after all cuts. Also shown is the statistical significance $S/\sqrt{B}$ with integrated luminosity of $\cal{L}$=300~$\mathrm{fb}^{-1}$ and $\cal{L}$=3000~$\mathrm{fb}^{-1}$. For illustration, we have used $m_{N_{1}}=30~\gev$ and $\Delta\phi=\pi/2$. }\label{table1}
  \centering
  \begin{tabular}{|l|c|c|c|c|}
  \hline
   & \multicolumn{2}{c|}{14 TeV} & \multicolumn{2}{c|}{100 TeV} \\
  \cline{2-5}
   & $\sigma(pp \to b\bar{b} l^\pm l^\pm 4j)$ [$\mathrm{fb}$]  & $\sigma(pp \rightarrow t\bar{t}W^{\pm}j)$ [$\mathrm{fb}$]
   & $\sigma(pp \to b\bar{b} l^\pm l^\pm 4j)$ [$\mathrm{fb}$]  & $\sigma(pp \rightarrow t\bar{t}W^{\pm}j)$ [$\mathrm{fb}$] \\
  \hline
  Cut-I & $1.21\times 10^{-2}$  & $1.81\times 10^{-1}$ & $3.22\times 10^{-1}$  & 3.01 \\
  \hline
  Cut-II & $7.54\times 10^{-3}$ & $8.48\times 10^{-3}$ & $1.98\times 10^{-1}$  & $2.99\times 10^{-2}$ \\
  \hline
  Cut-III & $4.35\times 10^{-3}$  & $3.16\times 10^{-4}$ & $1.15\times 10^{-1}$  & $2.82\times 10^{-3}$ \\
  \hline
  $S/\sqrt{B}$ with $\cal{L}$=300~$\mathrm{fb}^{-1}$ & \multicolumn{2}{c|}{4.24} & \multicolumn{2}{c|}{37.18} \\
  \hline
  $S/\sqrt{B}$ with $\cal{L}$=3000~$\mathrm{fb}^{-1}$ & \multicolumn{2}{c|}{13.40} & \multicolumn{2}{c|}{117.60} \\
  \hline
\end{tabular}
\end{table}

\begin{table}[!htbp]
  \caption{The value of $\overline{{\cal A}_{\rm CP}}$ at 14 TeV and 100 TeV LHC after all cuts for $m_{N_{1}}=30~\gev$ and $\Delta\phi=\pm\pi/4, \pm\pi/2, \pm4\pi/5$.}\label{table2}
  \centering
  \begin{tabular}{|l|c|c|}
  \hline
    &  ~~~~~14 TeV~~~~~ &  ~~~~~100 TeV~~~~~  \\
  \hline
  ~~~$\Delta\phi=-4\pi/5$~~~ &  ~~~0.67~~~ &   ~~~0.67~~~ \\
  \hline
  ~~~$\Delta\phi=-\pi/2$~~~  &  ~~~0.40~~~ &    ~~~0.40~~~ \\
  \hline
  ~~~$\Delta\phi=-\pi/4$~~~  &  ~~~0.18~~~ &    ~~~0.18~~~ \\
  \hline
  ~~~$\Delta\phi=\pi/4$~~~  &  ~~~-0.18~~~ &   ~~~ -0.18~~~ \\
  \hline
 ~~~$\Delta\phi=\pi/2$~~~  &  ~~~-0.40~~~ &    ~~~-0.40~~~ \\
  \hline
  ~~~$\Delta\phi=4\pi/5$~~~  & ~~~ -0.67~~~ &    ~~~-0.67~~~ \\
  \hline
\end{tabular}
\end{table}

\section{Summary}\label{sec5}

The existence of heavy Majorana neutrinos provides one of the most promising explanations for the origin of neutrino masses.
In this paper, we investigate the CP violation in top quark pair production and their rare lepton-number-violating decay at the LHC.
The significant interference of contributions from two nearly-degenerate Majorana neutrinos can lead to a CP-violating effect between $t \to b l_{}^{+}l_{}^{+}X$ and $\bar{t} \to \bar{b} l_{}^{-}  l_{}^{-} X$ processes.
It is found that in the Majorana neutrino mass range of $10~\gev < \mn < 80~\gev$, the CP asymmetry is independent of the Majorana neutrino mass at the LHC.
Taking $m_{N_{1}}=30~\gev$ and $\Delta\phi=\pi/2$ as an example, we explore the discovery prospects of CP violation at 14 TeV and 100 TeV LHC. We find that, the signal events are hard to be detected at 14 TeV LHC, while it is possible at 100 TeV LHC.
Furthermore, we investigate the CP asymmetry $\overline{{\cal A}_{\rm CP}}$ for $\Delta\phi = \pm\pi/4, \pm\pi/2$ and $\pm4\pi/5$ after the accepted cuts for  $m_{N_1}$ = 30 GeV at the LHC.
Once this kind of CP violation effects induced by the Majorana phase in top quark rare decay are observed at the LHC or future high energy colliders, they will be the clear evidence of new physics beyond the standard model.

\section*{Acknowledgements}
The authors thank the members of the Institute of theoretical physics of Shandong University for their helpful discussions. This work is supported by National Natural Science Foundation of China (grant Nos. 11875179, 11775130).

\begin{appendix}

\section{Calculation of the squared scattering amplitude }\label{appA}

In this appendix, we show explicitly the squared scattering amplitude given in Eq.~\ref{8.2}.
The functions ${\cal T}_i$ ($i = 1, 2$) and ${\cal T}_{12}$ can be respectively expressed as
\begin{align}
\label{A1}
{\cal T}_i =& \left|D_{N_i}\left(p_N^2\right)\right|^2 \cdot {\cal F} + \left|D_{N_i}\left({p_N^\prime}^2\right)\right|^2 \cdot {\cal G}  \nn \\
&- {\rm Re} \left[D_{N_i}\left(p_N^2\right)D_{N_i}^\ast\left({p_N^\prime}^2\right)\right] \cdot {\cal I}
- {\rm Im} \left[D_{N_i}\left(p_N^2\right)D_{N_i}^\ast\left({p_N^\prime}^2\right)\right] \cdot {\cal J} \; , \\
\label{A2}
{\cal T}_{12} =& 2 D_{N_1}\left(p_N^2\right)D_{N_2}^\ast\left({p_N}^2\right) \cdot {\cal F}
+ 2 D_{N_1}\left({p_N^\prime}^2\right)D_{N_2}^\ast\left({p_N^\prime}^2\right) \cdot {\cal G}   \nn \\
&- \left[D_{N_1}\left(p_N^2\right)D_{N_2}^\ast\left({p_N^\prime}^2\right) + D_{N_1}\left({p_N^\prime}^2\right)D_{N_2}^\ast\left(p_N^2\right)\right] \cdot {\cal I} \nn \\
&+ i \left[D_{N_1}\left(p_N^2\right)D_{N_2}^\ast\left({p_N^\prime}^2\right) - D_{N_1}\left({p_N^\prime}^2\right)D_{N_2}^\ast\left(p_N^2\right)\right] \cdot {\cal J} \; ,
\end{align}
where $p_N=p_1-p_2-p_3$ and $p_N^\prime=p_1-p_2-p_4$. $D_X\left(p^2\right)$ is the Breit-Wigner propagator and can be defined as
\begin{align}
\label{A3}
D_X\left(p^2\right) = \frac{1}{p^2-m_X^2+im_X\Gamma_X} \; ,
\end{align}
with $m_X$ and $\Gamma_X$ being the mass and total decay width of the corresponding particles.

The explicit expressions of ${\cal F}$, ${\cal G}$, ${\cal I}$ and ${\cal J}$ introduced in Eq.~\ref{A1} and Eq.~\ref{A2} can be given by
\begin{align}
\label{}
{\cal F} = &\left\{ -2 m_t^2 m_W^2 (p_1 \cdot p_5) (p_2 \cdot p_3) - 2 m_W^2 \left(m_t^2-2 m_W^2\right) (p_1 \cdot p_3) (p_2 \cdot p_5) + m_t^2\biggl[4 m_W^2 (p_2 \cdot p_3) (p_2 \cdot p_5)\right. \biggr. \nn \\
&+ \left.\biggl.(p_1 \cdot p_2) \left[\left(-m_t^2+2 m_W^2+2 (p_1 \cdot p_2)\right) (p_3 \cdot p_5)+2 (p_3 \cdot p_w) (p_5 \cdot p_w)\right]\biggr]\right\}(p_4 \cdot p_6) \; , \\
{\cal G} = &\left\{ -2 m_t^2 m_W^2 (p_1 \cdot p_5) (p_2 \cdot p_4) - 2 m_W^2 \left(m_t^2-2 m_W^2\right) (p_1 \cdot p_4) (p_2 \cdot p_5) + m_t^2\biggl[4 m_W^2 (p_2 \cdot p_4) (p_2 \cdot p_5)\right. \biggr.  \nn \\
&+ \left.\biggl.(p_1 \cdot p_2) \left[\left(-m_t^2+2 m_W^2+2 (p_1 \cdot p_2)\right) (p_4 \cdot p_5)+2 (p_4 \cdot p_w) (p_5 \cdot p_w)\right]\biggr]\right\}(p_3 \cdot p_6) \; , \\
{\cal I} = & 2 m_W^2 \biggl\{\left(m_t^2 -2 m_W^2\right) (p_2 \cdot p_5)\left[(p_1 \cdot p_3) (p_4 \cdot p_6)+ (p_1 \cdot p_4) (p_3 \cdot p_6) - (p_1 \cdot p_6) (p_3 \cdot p_4)\right] \biggr. \nn \\
&\biggl.- m_t^2 \left[(p_2 \cdot p_5)-(p_5 \cdot p_w)\right] \left[(p_2 \cdot p_3) (p_4 \cdot p_6)+ (p_2 \cdot p_4) (p_3 \cdot p_6) - (p_2 \cdot p_6) (p_3 \cdot p_4)\right]\biggr\} \nn \\
&+m_t^2 \left(m_t^2 -2 m_W^2-2(p_1 \cdot p_2)\right) (p_1 \cdot p_2)\left[(p_3 \cdot p_6) (p_4 \cdot p_5)+ (p_3 \cdot p_5) (p_4 \cdot p_6) - (p_3 \cdot p_4) (p_5 \cdot p_6)\right]   \nn \\
&+2m_t^2(p_1 \cdot p_2)(p_5 \cdot p_w)\left[(p_3 \cdot p_4) (p_6 \cdot p_w) - (p_3 \cdot p_6) (p_4 \cdot p_w) - (p_4 \cdot p_6) (p_3 \cdot p_w)\right] \; ,  \\
{\cal J} = &
2 m_W^2 \biggl\{2 m_W^2 (p_2 \cdot p_5) - m_t^2 \left[(p_1 \cdot p_5) - (p_2 \cdot p_5) \right]\biggr\}\epsilon_{p_1 p_2 p_3 p_4}
+ \biggl\{4 m_W^4 (p_2 \cdot p_5) - m_t^4(p_1 \cdot p_2) \biggr. \nn \\
&+ \biggl.2 m_t^2\left[(p_1 \cdot p_2)(m_W^2+(p_1 \cdot p_2)+(p_1 \cdot p_5))-(m_W^2+(p_1 \cdot p_2))(p_2 \cdot p_5)\right]\biggr\}\epsilon_{p_1 p_3 p_4 p_5} \nn \\
&+ m_t^2\biggl\{m_t^2(p_1 \cdot p_2)-2\left[m_W^2+(p_1 \cdot p_2)\right]\left[(p_1 \cdot p_2)+(p_1 \cdot p_5)\right]+2\left[m_W^2+(p_1 \cdot p_2)\right](p_2 \cdot p_5)\biggr\}\epsilon_{p_2 p_3 p_4 p_5} \; ,
\end{align}
where $\epsilon_{p_1 p_2 p_3 p_4}=\epsilon_{\mu \nu \rho \sigma}p_1^\mu p_2^\nu p_3^\rho p_4^\sigma$.

\end{appendix}

\end{document}